\documentclass[12pt]{article}
\usepackage{a4,epsfig,amssymb}
\newcommand{\rb}[1]{\raisebox{1.5ex}[-1.5ex]{#1}}

\newcommand{\PZz}{\ensuremath{\mathrm{Z}}}

\newcommand{\PSnu}{\ensuremath{\widetilde{\nu}}}

\newcommand{\Pnu}{\ensuremath{\nu}}
\newcommand{\Pe}{\ensuremath{\mathrm{e}}}
\newcommand{\Panu}{\ensuremath{\overline{\nu}}}
\newcommand{\PSl}{\ensuremath{\widetilde{\ell}}}
\newcommand{\PSe}{\ensuremath{\widetilde{\mathrm{e}}}}
\newcommand{\PSmu}{\ensuremath{\widetilde{\mu}}}
\newcommand{\PStau}{\ensuremath{\widetilde{\tau}}}

\newcommand{\Pchi}{\ensuremath{\chi}}

\newcommand{\Gc}{\ensuremath{\mathrm{GeV}/c}}
\newcommand{\Gcsq}{\ensuremath{\mathrm{GeV}/c^2}}
\newcommand{\Tcsq}{\ensuremath{\mathrm{TeV}/c^2}}
\newcommand{\tb}{\ensuremath{\tan\beta}}
\newcommand{\sqtw}{\ensuremath{\sin^2\vartheta_W}}
\newcommand{\pbinv}{\ensuremath{\mathrm{pb}^{-1}}}
\newcommand{\epem}{\ensuremath{\mathrm{e^+e^-}}}
\newcommand{\gaga}{\ensuremath{\gamma\gamma}}
\newcommand{\pT}{\ensuremath{p_{\mathrm{T}}}}
\newcommand{\wwchi}{\ensuremath{\chi_{\mathrm{WW}}}}
\newcommand{\tchannel}{\ensuremath{t} channel}
\newcommand{\schannel}{\ensuremath{s} channel}
\newcommand{\nbarnf}{\ensuremath{\bar{N}_{95}}}
\newcommand{\dum}{\phantom{0}}
\newcommand{\pdum}{\phantom{.0}}
\begin{document}

\begin{titlepage}

{\Large\centerline{EUROPEAN LABORATORY FOR PARTICLE PHYSICS (CERN)}}
\vspace{1cm}
\begin{flushright}
CERN-PPE/97-056 \\
27 May 1997
\end{flushright}
\begin{center}
\vspace{3cm}
{\huge\bf Search for sleptons in {\boldmath \epem{}} collisions at 
centre--of--mass energies of 161~GeV and 172~GeV}\\
{\Large \vspace{7ex} The ALEPH Collaboration}
\end{center}
\vspace{1cm}
\begin{center}
{\bf Abstract}
\end{center}

The data recorded by the ALEPH experiment at LEP at 
centre--of--mass energies of 
161~GeV and 172~GeV were analysed to search for sleptons, the 
supersymmetric partners of leptons. 
No evidence for the production of these particles was found. 
The number of candidates observed is consistent with the 
background expected from four--fermion processes and \gaga{}--interactions.
Improved mass limits at 95\% C.L. are reported.
\\[5cm]
\centerline{\em (Submitted to Physics Letters B)}
\end{titlepage}
\pagestyle{empty}
\newpage
\small
%
%
\newlength{\saveparskip}
\newlength{\savetextheight}
\newlength{\savetopmargin}
\newlength{\savetextwidth}
\newlength{\saveoddsidemargin}
\newlength{\savetopsep}
\setlength{\saveparskip}{\parskip}
\setlength{\savetextheight}{\textheight}
\setlength{\savetopmargin}{\topmargin}
\setlength{\savetextwidth}{\textwidth}
\setlength{\saveoddsidemargin}{\oddsidemargin}
\setlength{\savetopsep}{\topsep}
%
%
\setlength{\parskip}{0.0cm}
\setlength{\textheight}{24.0cm}
\setlength{\topmargin}{-1.5cm}
\setlength{\textwidth}{16 cm}
\setlength{\oddsidemargin}{-0.0cm}
\setlength{\topsep}{1mm}
\pretolerance=10000
\centerline{\large\bf The ALEPH Collaboration}
\footnotesize
\vspace{0.5cm}
{\raggedbottom
\begin{sloppypar}
\samepage\noindent
R.~Barate,
D.~Buskulic,
D.~Decamp,
P.~Ghez,
C.~Goy,
J.-P.~Lees,
A.~Lucotte,
M.-N.~Minard,
J.-Y.~Nief,
B.~Pietrzyk
\nopagebreak
\begin{center}
\parbox{15.5cm}{\sl\samepage
Laboratoire de Physique des Particules (LAPP), IN$^{2}$P$^{3}$-CNRS,
74019 Annecy-le-Vieux Cedex, France}
\end{center}\end{sloppypar}
\vspace{2mm}
\begin{sloppypar}
\noindent
M.P.~Casado,
M.~Chmeissani,
P.~Comas,
J.M.~Crespo,
M.~Delfino, 
E.~Fernandez,
M.~Fernandez-Bosman,
Ll.~Garrido,$^{15}$
A.~Juste,
M.~Martinez,
R.~Miquel,
Ll.M.~Mir,
S.~Orteu,
C.~Padilla,
I.C.~Park,
A.~Pascual,
J.A.~Perlas,
I.~Riu,
F.~Sanchez,
F.~Teubert
\nopagebreak
\begin{center}
\parbox{15.5cm}{\sl\samepage
Institut de F\'{i}sica d'Altes Energies, Universitat Aut\`{o}noma
de Barcelona, 08193 Bellaterra (Barcelona), Spain$^{7}$}
\end{center}\end{sloppypar}
\vspace{2mm}
\begin{sloppypar}
\noindent
A.~Colaleo,
D.~Creanza,
M.~de~Palma,
G.~Gelao,
G.~Iaselli,
G.~Maggi,
M.~Maggi,
N.~Marinelli,
S.~Nuzzo,
A.~Ranieri,
G.~Raso,
F.~Ruggieri,
G.~Selvaggi,
L.~Silvestris,
P.~Tempesta,
A.~Tricomi,$^{3}$
G.~Zito
\nopagebreak
\begin{center}
\parbox{15.5cm}{\sl\samepage
Dipartimento di Fisica, INFN Sezione di Bari, 70126
Bari, Italy}
\end{center}\end{sloppypar}
\vspace{2mm}
\begin{sloppypar}
\noindent
X.~Huang,
J.~Lin,
Q. Ouyang,
T.~Wang,
Y.~Xie,
R.~Xu,
S.~Xue,
J.~Zhang,
L.~Zhang,
W.~Zhao
\nopagebreak
\begin{center}
\parbox{15.5cm}{\sl\samepage
Institute of High-Energy Physics, Academia Sinica, Beijing, The People's
Republic of China$^{8}$}
\end{center}\end{sloppypar}
\vspace{2mm}
\begin{sloppypar}
\noindent
D.~Abbaneo,
R.~Alemany,
A.O.~Bazarko,$^{1}$
U.~Becker,
P.~Bright-Thomas,
M.~Cattaneo,
F.~Cerutti,
G.~Dissertori,
H.~Drevermann,
R.W.~Forty,
M.~Frank,
R.~Hagelberg,
J.B.~Hansen,
J.~Harvey,
P.~Janot,
B.~Jost,
E.~Kneringer,
J.~Knobloch,
I.~Lehraus,
G.~Lutters,
P.~Mato,
A.~Minten,
L.~Moneta,
A.~Pacheco,
J.-F.~Pusztaszeri,$^{20}$
F.~Ranjard,
G.~Rizzo,
L.~Rolandi,
D.~Rousseau,
D.~Schlatter,
M.~Schmitt,
O.~Schneider,
W.~Tejessy,
I.R.~Tomalin,
H.~Wachsmuth,
A.~Wagner$^{21}$
\nopagebreak
\begin{center}
\parbox{15.5cm}{\sl\samepage
European Laboratory for Particle Physics (CERN), 1211 Geneva 23,
Switzerland}
\end{center}\end{sloppypar}
\vspace{2mm}
\begin{sloppypar}
\noindent
Z.~Ajaltouni,
A.~Barr\`{e}s,
C.~Boyer,
A.~Falvard,
C.~Ferdi,
P.~Gay,
C~.~Guicheney,
P.~Henrard,
J.~Jousset,
B.~Michel,
S.~Monteil,
J-C.~Montret,
D.~Pallin,
P.~Perret,
F.~Podlyski,
J.~Proriol,
P.~Rosnet,
J.-M.~Rossignol
\nopagebreak
\begin{center}
\parbox{15.5cm}{\sl\samepage
Laboratoire de Physique Corpusculaire, Universit\'e Blaise Pascal,
IN$^{2}$P$^{3}$-CNRS, Clermont-Ferrand, 63177 Aubi\`{e}re, France}
\end{center}\end{sloppypar}
\vspace{2mm}
\begin{sloppypar}
\noindent
T.~Fearnley,
J.D.~Hansen,
J.R.~Hansen,
P.H.~Hansen,
B.S.~Nilsson,
B.~Rensch,
A.~W\"a\"an\"anen
\begin{center}
\parbox{15.5cm}{\sl\samepage
Niels Bohr Institute, 2100 Copenhagen, Denmark$^{9}$}
\end{center}\end{sloppypar}
\vspace{2mm}
\begin{sloppypar}
\noindent
G.~Daskalakis,
A.~Kyriakis,
C.~Markou,
E.~Simopoulou,
A.~Vayaki
\nopagebreak
\begin{center}
\parbox{15.5cm}{\sl\samepage
Nuclear Research Center Demokritos (NRCD), Athens, Greece}
\end{center}\end{sloppypar}
\vspace{2mm}
\begin{sloppypar}
\noindent
A.~Blondel,
J.C.~Brient,
F.~Machefert,
A.~Roug\'{e},
M.~Rumpf,
A.~Valassi,$^{6}$
H.~Videau
\nopagebreak
\begin{center}
\parbox{15.5cm}{\sl\samepage
Laboratoire de Physique Nucl\'eaire et des Hautes Energies, Ecole
Polytechnique, IN$^{2}$P$^{3}$-CNRS, 91128 Palaiseau Cedex, France}
\end{center}\end{sloppypar}
\vspace{2mm}
\begin{sloppypar}
\noindent
E.~Focardi,
G.~Parrini,
K.~Zachariadou
\nopagebreak
\begin{center}
\parbox{15.5cm}{\sl\samepage
Dipartimento di Fisica, Universit\`a di Firenze, INFN Sezione di Firenze,
50125 Firenze, Italy}
\end{center}\end{sloppypar}
\vspace{2mm}
\begin{sloppypar}
\noindent
R.~Cavanaugh,
M.~Corden,
C.~Georgiopoulos,
T.~Huehn,
D.E.~Jaffe
\nopagebreak
\begin{center}
\parbox{15.5cm}{\sl\samepage
Supercomputer Computations Research Institute,
Florida State University,
Tallahassee, FL 32306-4052, USA $^{13,14}$}
\end{center}\end{sloppypar}
\vspace{2mm}
\begin{sloppypar}
\noindent
A.~Antonelli,
G.~Bencivenni,
G.~Bologna,$^{4}$
F.~Bossi,
P.~Campana,
G.~Capon,
D.~Casper,
V.~Chiarella,
G.~Felici,
P.~Laurelli,
G.~Mannocchi,$^{5}$
F.~Murtas,
G.P.~Murtas,
L.~Passalacqua,
M.~Pepe-Altarelli
\nopagebreak
\begin{center}
\parbox{15.5cm}{\sl\samepage
Laboratori Nazionali dell'INFN (LNF-INFN), 00044 Frascati, Italy}
\end{center}\end{sloppypar}
\vspace{2mm}
\pagebreak
\begin{sloppypar}
\noindent
L.~Curtis,
S.J.~Dorris,
A.W.~Halley,
I.G.~Knowles,
J.G.~Lynch,
V.~O'Shea,
C.~Raine,
J.M.~Scarr,
K.~Smith,
P.~Teixeira-Dias,
A.S.~Thompson,
E.~Thomson,
F.~Thomson,
R.M.~Turnbull
\nopagebreak
\begin{center}
\parbox{15.5cm}{\sl\samepage
Department of Physics and Astronomy, University of Glasgow, Glasgow G12
8QQ,United Kingdom$^{10}$}
\end{center}\end{sloppypar}
\vspace{2mm}
\begin{sloppypar}
\noindent
C.~Geweniger,
G.~Graefe,
P.~Hanke,
G.~Hansper,
V.~Hepp,
E.E.~Kluge,
A.~Putzer,
M.~Schmidt,
J.~Sommer,
K.~Tittel,
S.~Werner,
M.~Wunsch
\begin{center}
\parbox{15.5cm}{\sl\samepage
Institut f\"ur Hochenergiephysik, Universit\"at Heidelberg, 69120
Heidelberg, Fed.\ Rep.\ of Germany$^{16}$}
\end{center}\end{sloppypar}
\vspace{2mm}
\begin{sloppypar}
\noindent
R.~Beuselinck,
D.M.~Binnie,
W.~Cameron,
P.J.~Dornan,
M.~Girone,
S.~Goodsir,
E.B.~Martin,
P.~Morawitz,
A.~Moutoussi,
J.~Nash,
J.K.~Sedgbeer,
P.~Spagnolo,
A.M.~Stacey,
M.D.~Williams
\nopagebreak
\begin{center}
\parbox{15.5cm}{\sl\samepage
Department of Physics, Imperial College, London SW7 2BZ,
United Kingdom$^{10}$}
\end{center}\end{sloppypar}
\vspace{2mm}
\begin{sloppypar}
\noindent
V.M.~Ghete,
P.~Girtler,
D.~Kuhn,
G.~Rudolph
\nopagebreak
\begin{center}
\parbox{15.5cm}{\sl\samepage
Institut f\"ur Experimentalphysik, Universit\"at Innsbruck, 6020
Innsbruck, Austria$^{18}$}
\end{center}\end{sloppypar}
\vspace{2mm}
\begin{sloppypar}
\noindent
A.P.~Betteridge,
C.K.~Bowdery,
P.~Colrain,
G.~Crawford,
A.J.~Finch,
F.~Foster,
G.~Hughes,
R.W.~Jones,
T.~Sloan,
E.P.~Whelan,
M.I.~Williams
\nopagebreak
\begin{center}
\parbox{15.5cm}{\sl\samepage
Department of Physics, University of Lancaster, Lancaster LA1 4YB,
United Kingdom$^{10}$}
\end{center}\end{sloppypar}
\vspace{2mm}
\begin{sloppypar}
\noindent
C.~Hoffmann,
K.~Jakobs,
K.~Kleinknecht,
G.~Quast,
B.~Renk,
E.~Rohne,
H.-G.~Sander,
P.~van~Gemmeren,
C.~Zeitnitz
\nopagebreak
\begin{center}
\parbox{15.5cm}{\sl\samepage
Institut f\"ur Physik, Universit\"at Mainz, 55099 Mainz, Fed.\ Rep.\
of Germany$^{16}$}
\end{center}\end{sloppypar}
\vspace{2mm}
\begin{sloppypar}
\noindent
J.J.~Aubert,
C.~Benchouk,
A.~Bonissent,
G.~Bujosa,
D.~Calvet,
J.~Carr,
P.~Coyle,
C.~Diaconu,
N.~Konstantinidis,
O.~Leroy,
F.~Motsch,
P.~Payre,
M.~Talby,
A.~Sadouki,
M.~Thulasidas,
A.~Tilquin,
K.~Trabelsi
\nopagebreak
\begin{center}
\parbox{15.5cm}{\sl\samepage
Centre de Physique des Particules, Facult\'e des Sciences de Luminy,
IN$^{2}$P$^{3}$-CNRS, 13288 Marseille, France}
\end{center}\end{sloppypar}
\vspace{2mm}
\begin{sloppypar}
\noindent
M.~Aleppo, 
F.~Ragusa$^{12}$
\nopagebreak
\begin{center}
\parbox{15.5cm}{\sl\samepage
Dipartimento di Fisica, Universit\`a di Milano e INFN Sezione di
Milano, 20133 Milano, Italy.}
\end{center}\end{sloppypar}
\vspace{2mm}
\begin{sloppypar}
\noindent
R.~Berlich,
W.~Blum,
V.~B\"uscher,
H.~Dietl,
G.~Ganis,
C.~Gotzhein,
H.~Kroha,
G.~L\"utjens,
G.~Lutz,
W.~M\"anner,
H.-G.~Moser,
R.~Richter,
A.~Rosado-Schlosser,
S.~Schael,
R.~Settles,
H.~Seywerd,
R.~St.~Denis,
H.~Stenzel,
W.~Wiedenmann,
G.~Wolf
\nopagebreak
\begin{center}
\parbox{15.5cm}{\sl\samepage
Max-Planck-Institut f\"ur Physik, Werner-Heisenberg-Institut,
80805 M\"unchen, Fed.\ Rep.\ of Germany\footnotemark[16]}
\end{center}\end{sloppypar}
\vspace{2mm}
\begin{sloppypar}
\noindent
J.~Boucrot,
O.~Callot,$^{12}$
S.~Chen,
A.~Cordier,
M.~Davier,
L.~Duflot,
J.-F.~Grivaz,
Ph.~Heusse,
A.~H\"ocker,
A.~Jacholkowska,
M.~Jacquet,
D.W.~Kim,$^{2}$
F.~Le~Diberder,
J.~Lefran\c{c}ois,
A.-M.~Lutz,
I.~Nikolic,
M.-H.~Schune,
L.~Serin,
S.~Simion,
E.~Tournefier,
J.-J.~Veillet,
I.~Videau,
D.~Zerwas
\nopagebreak
\begin{center}
\parbox{15.5cm}{\sl\samepage
Laboratoire de l'Acc\'el\'erateur Lin\'eaire, Universit\'e de Paris-Sud,
IN$^{2}$P$^{3}$-CNRS, 91405 Orsay Cedex, France}
\end{center}\end{sloppypar}
\vspace{2mm}
\begin{sloppypar}
\noindent
\samepage
P.~Azzurri,
G.~Bagliesi,
S.~Bettarini,
C.~Bozzi,
G.~Calderini,
V.~Ciulli,
R.~Dell'Orso,
R.~Fantechi,
I.~Ferrante,
A.~Giassi,
A.~Gregorio,
F.~Ligabue,
A.~Lusiani,
P.S.~Marrocchesi,
A.~Messineo,
F.~Palla,
G.~Sanguinetti,
A.~Sciab\`a,
J.~Steinberger,
R.~Tenchini,
C.~Vannini,
A.~Venturi,
P.G.~Verdini
\samepage
\begin{center}
\parbox{15.5cm}{\sl\samepage
Dipartimento di Fisica dell'Universit\`a, INFN Sezione di Pisa,
e Scuola Normale Superiore, 56010 Pisa, Italy}
\end{center}\end{sloppypar}
\vspace{2mm}
\begin{sloppypar}
\noindent
G.A.~Blair,
L.M.~Bryant,
J.T.~Chambers,
Y.~Gao,
M.G.~Green,
T.~Medcalf,
P.~Perrodo,
J.A.~Strong,
J.H.~von~Wimmersperg-Toeller
\nopagebreak
\begin{center}
\parbox{15.5cm}{\sl\samepage
Department of Physics, Royal Holloway \& Bedford New College,
University of London, Surrey TW20 OEX, United Kingdom$^{10}$}
\end{center}\end{sloppypar}
\vspace{2mm}
\begin{sloppypar}
\noindent
D.R.~Botterill,
R.W.~Clifft,
T.R.~Edgecock,
S.~Haywood,
P.~Maley,
P.R.~Norton,
J.C.~Thompson,
A.E.~Wright
\nopagebreak
\begin{center}
\parbox{15.5cm}{\sl\samepage
Particle Physics Dept., Rutherford Appleton Laboratory,
Chilton, Didcot, Oxon OX11 OQX, United Kingdom$^{10}$}
\end{center}\end{sloppypar}
\pagebreak
\vspace{2mm}
\begin{sloppypar}
\noindent
B.~Bloch-Devaux,
P.~Colas,
B.~Fabbro,
W.~Kozanecki,
E.~Lan\c{c}on,
M.C.~Lemaire,
E.~Locci,
P.~Perez,
J.~Rander,
J.-F.~Renardy,
A.~Rosowsky,
A.~Roussarie,
J.-P.~Schuller,
J.~Schwindling,
A.~Trabelsi,
B.~Vallage
\nopagebreak
\begin{center}
\parbox{15.5cm}{\sl\samepage
CEA, DAPNIA/Service de Physique des Particules,
CE-Saclay, 91191 Gif-sur-Yvette Cedex, France$^{17}$}
\end{center}\end{sloppypar}
\vspace{2mm}
\begin{sloppypar}
\noindent
S.N.~Black,
J.H.~Dann,
H.Y.~Kim,
A.M.~Litke,
M.A. McNeil,
G.~Taylor
\nopagebreak
\begin{center}
\parbox{15.5cm}{\sl\samepage
Institute for Particle Physics, University of California at
Santa Cruz, Santa Cruz, CA 95064, USA$^{19}$}
\end{center}\end{sloppypar}
\vspace{2mm}
\begin{sloppypar}
\noindent
C.N.~Booth,
R.~Boswell,
C.A.J.~Brew,
S.~Cartwright,
F.~Combley,
M.S.~Kelly,
M.~Lehto,
W.M.~Newton,
J.~Reeve,
L.F.~Thompson
\nopagebreak
\begin{center}
\parbox{15.5cm}{\sl\samepage
Department of Physics, University of Sheffield, Sheffield S3 7RH,
United Kingdom$^{10}$}
\end{center}\end{sloppypar}
\vspace{2mm}
\begin{sloppypar}
\noindent
K.~Affholderbach,
A.~B\"ohrer,
S.~Brandt,
G.~Cowan,
J.~Foss,
C.~Grupen,
P.~Saraiva,
L.~Smolik,
F.~Stephan 
\nopagebreak
\begin{center}
\parbox{15.5cm}{\sl\samepage
Fachbereich Physik, Universit\"at Siegen, 57068 Siegen,
 Fed.\ Rep.\ of Germany$^{16}$}
\end{center}\end{sloppypar}
\vspace{2mm}
\begin{sloppypar}
\noindent
M.~Apollonio,
L.~Bosisio,
R.~Della~Marina,
G.~Giannini,
B.~Gobbo,
G.~Musolino
\nopagebreak
\begin{center}
\parbox{15.5cm}{\sl\samepage
Dipartimento di Fisica, Universit\`a di Trieste e INFN Sezione di Trieste,
34127 Trieste, Italy}
\end{center}\end{sloppypar}
\vspace{2mm}
\begin{sloppypar}
\noindent
J.~Putz,
J.~Rothberg,
S.~Wasserbaech,
R.W.~Williams
\nopagebreak
\begin{center}
\parbox{15.5cm}{\sl\samepage
Experimental Elementary Particle Physics, University of Washington, WA 98195
Seattle, U.S.A.}
\end{center}\end{sloppypar}
\vspace{2mm}
\begin{sloppypar}
\noindent
S.R.~Armstrong,
E.~Charles,
P.~Elmer,
D.P.S.~Ferguson,
S.~Gonz\'{a}lez,
T.C.~Greening,
O.J.~Hayes,
H.~Hu,
S.~Jin,
P.A.~McNamara III,
J.M.~Nachtman,
J.~Nielsen,
W.~Orejudos,
Y.B.~Pan,
Y.~Saadi,
I.J.~Scott,
J.~Walsh,
Sau~Lan~Wu,
X.~Wu,
J.M.~Yamartino,
G.~Zobernig
\nopagebreak
\begin{center}
\parbox{15.5cm}{\sl\samepage
Department of Physics, University of Wisconsin, Madison, WI 53706,
USA$^{11}$}
\end{center}\end{sloppypar}
}
\footnotetext[1]{Now at Princeton University, Princeton, NJ 08544, U.S.A.}
\footnotetext[2]{Permanent address: Kangnung National University, Kangnung,
Korea.}
\footnotetext[3]{Also at Dipartimento di Fisica, INFN Sezione di Catania,
Catania, Italy.}
\footnotetext[4]{Also Istituto di Fisica Generale, Universit\`{a} di
Torino, Torino, Italy.}
\footnotetext[5]{Also Istituto di Cosmo-Geofisica del C.N.R., Torino,
Italy.}
\footnotetext[6]{Supported by the Commission of the European Communities,
contract ERBCHBICT941234.}
\footnotetext[7]{Supported by CICYT, Spain.}
\footnotetext[8]{Supported by the National Science Foundation of China.}
\footnotetext[9]{Supported by the Danish Natural Science Research Council.}
\footnotetext[10]{Supported by the UK Particle Physics and Astronomy Research
Council.}
\footnotetext[11]{Supported by the US Department of Energy, grant
DE-FG0295-ER40896.}
\footnotetext[12]{Also at CERN, 1211 Geneva 23,Switzerland.}
\footnotetext[13]{Supported by the US Department of Energy, contract
DE-FG05-92ER40742.}
\footnotetext[14]{Supported by the US Department of Energy, contract
DE-FC05-85ER250000.}
\footnotetext[15]{Permanent address: Universitat de Barcelona, 08208 Barcelona,
Spain.}
\footnotetext[16]{Supported by the Bundesministerium f\"ur Bildung,
Wissenschaft, Forschung und Technologie, Fed. Rep. of Germany.}
\footnotetext[17]{Supported by the Direction des Sciences de la
Mati\`ere, C.E.A.}
\footnotetext[18]{Supported by Fonds zur F\"orderung der wissenschaftlichen
Forschung, Austria.}
\footnotetext[19]{Supported by the US Department of Energy,
grant DE-FG03-92ER40689.}
\footnotetext[20]{Now at School of Operations Research and Industrial
Engireering, Cornell University, Ithaca, NY 14853-3801, U.S.A.}
\footnotetext[21]{Now at Schweizerischer Bankverein, Basel, Switzerland.}
%
%
\setlength{\parskip}{\saveparskip}
\setlength{\textheight}{\savetextheight}
\setlength{\topmargin}{\savetopmargin}
\setlength{\textwidth}{\savetextwidth}
\setlength{\oddsidemargin}{\saveoddsidemargin}
\setlength{\topsep}{\savetopsep}
\normalsize
\newpage
\pagestyle{plain}
\setcounter{page}{1}

\section{Introduction}

The main consequence of supersymmetric theories~\cite{jacob} is the doubling
of the particle spectrum: for each fermion's chirality state a scalar particle
is introduced. Depending on the chirality state they are associated to, the
scalar partners of the leptons (sleptons) are called ``right'' ($\PSl_R$) or
``left'' ($\PSl_L$). These are the eigenstates of the weak interaction.
Particle masses are generated by the Higgs mechanism with two doublets. 
The partners of the
Higgs and gauge bosons are the Higgsinos and gauginos. 
Exact
supersymmetry implies mass degeneracy for the particles and their
supersymmetric partners; since no evidence for supersymmetry has been
observed up to now, supersymmetry has to be a broken symmetry. 
When supersymmetry is broken, the weak eigenstates mix to form the 
mass eigenstates: 
neutral Higgsinos and gauginos mix to form the mass eigenstates called
``neutralinos'', and the charged Higgsinos and gauginos mix to form the
eigenstates called ``charginos''.
\par
In order to preserve one of the most appealing aspects of supersymmetry, i.e.,
possibly being a solution to the hierarchy problem, supersymmetric particles
must have masses of order~\Tcsq{} or less.
The increase of the centre--of--mass energy of LEP, CERN's large \epem{}
collider, to 140~GeV in 1995 and 
above the W pair production threshold in 1996 has opened a new energy
regime to be probed in search for 
supersymmetry~\cite{opal,all1.5,bib:lep15paper}.
In this letter the results of a search for the scalar partners of the 
leptons at centre--of--mass energies from 161~GeV to 172~GeV in the data 
recorded by the ALEPH detector in 1996 are presented. 
\par
The minimal supersymmetric extension of the standard model (MSSM)~\cite{jacob}
is used as a reference model. R--parity, a quantum number that distinguishes
standard model particles from supersymmetric particles \cite{rparity}, is
assumed to be conserved, implying that sleptons can only be produced in pairs.
In general the sleptons decay 
to their standard model partner
and the lightest neutralino ($\chi$) with an undetectable lifetime. 
The latter is assumed to be stable and 
escapes the apparatus undetected, leading to a final state of acoplanar
lepton pairs. In the following, any deviation from this behaviour will
be mentioned explicitly.
\par
The only dependence
of the cross section for smuon (\PSmu) and stau (\PStau) production
on supersymmetric parameters is through the slepton mass matrix.
The production proceeds via \schannel{} only, whereas 
selectrons (\PSe) can also be produced by exchanging neutralinos in the
\tchannel{}. 
The selectron cross section \cite{selectron} therefore  
depends on the selectron mass and, 
via the \tchannel{}, on
the MSSM parameters (the supersymmetric Higgs mass term $\mu$, the soft
supersymmetry breaking term associated to the $SU(2)_L$ group $M_2$, and the
ratio of the vacuum expectation values of the two Higgs doublets, \tb). 
\par
The off-diagonal elements of a slepton mass matrix are proportional to
the lepton's mass, therefore
left and right sleptons can mix to form the mass eigenstates. However,
mixing is expected to be negligible for smuons and selectrons due to the 
small masses of their standard model partners.
\par
The unification condition
$M_1 = \frac{5}{3}\tan^2{\vartheta_W}M_2$, where $M_1$ is the soft
supersymmetry breaking parameter associated with the $U(1)_Y$ group, is
assumed to be valid. 
In the following, the parameter space where $|\mu| \gg M_2$ ($|\mu| \ll M_2$)
will be referred to as the ``gaugino'' (``higgsino'') region, 
as suggested by the
field content of the lightest neutralino.

\subsection{The ALEPH Detector}

The ALEPH detector is described in detail in
Ref.~\cite{bib:detectorpaper}. An account of the performance of the detector
and a description of the standard analysis algorithms can be found in
Ref.~\cite{bib:performancepaper}. Here, only a brief description of the
detector elements and the algorithms relevant for this analysis is given.

In ALEPH, the trajectories of charged particles are measured with a silicon
vertex detector, a cylindrical drift chamber, and a large time
projection chamber (TPC). These detector components are located in a 
1.5~T magnetic field provided
by a superconducting solenoidal coil.
The electromagnetic calorimeter (ECAL), placed between the TPC and the coil, 
is a highly segmented sampling calorimeter which is used to identify electrons
and photons and to measure their energy and position.
The luminosity monitors (LCAL and SICAL) extend the calorimetric coverage 
down to 30~mrad from the beam axis, taking into account the 
additional shielding against beam related background, installed prior
to the 1996 running.
The hadron calorimeter (HCAL) consists of the iron return yoke of the magnet 
instrumented with streamer tubes. It provides a measurement of hadronic energy 
and, together with the external muon chambers, muon identification.

Global event quantities are measured with an energy flow algorithm. This
algorithm combines individual calorimeter and tracker measurements into
energy flow ``objects''. These objects are classified as photons, neutral
hadrons, and charged particles.

The present analysis makes use of lepton identification. 
In ALEPH, electrons are identified by the longitudinal and the transverse
energy distribution of the ECAL shower, and by the
specific ionization information in the TPC when 
available~\cite{bib:performancepaper}. Muons are
identified through their hit pattern in the HCAL and associated hits in the 
muon chambers.

\subsection{Data Sample}
For this analysis, the data taken in 1996 at centre--of--mass energies
of $161.3$, $170.3$ and $172.3$~GeV are used, 
which correspond to integrated
luminosities of 11.1~\pbinv{}, 1.1~\pbinv{} and 9.5~\pbinv{} respectively. 
In the following, these three centre--of--mass
energies will be referred to as 161 and 172~GeV, combining the two points
at 170.3 and 172.3~GeV.
Since the expected limits for sleptons are still in the kinematic reach of the 
5.7~\pbinv{} data taken at centre--of--mass energies of 130 and 136~GeV 
in 1995, the results published in~\cite{bib:lep15paper} are combined with the
results of the present analysis for the 161 and 172~GeV data.

\subsection{Monte Carlo Sample}
Samples of all the major background processes corresponding to at least
20 times the collected luminosity have been generated.
The most important background sources are lepton pair production, \gaga{}
collisions with lepton production, W pair production, \PZz$\gamma^{\ast}$
production (where $\gamma^{\ast}$ denotes a virtual \PZz{} or photon), \PZz{}ee
and We$\nu$ production. Bhabha processes were simulated with
UNIBAB~\cite{bib:unibab}, muon and tau pair production with
KORALZ~\cite{bib:koralz}, \gaga{} processes with PHOT02~\cite{bib:phot}
($\gaga\to$ leptons) and PYTHIA~\cite{bib:pythia} ($\gaga\to$ hadrons), WW
production with KORALW~\cite{bib:koralw}, and the remaining four--fermion
processes with PYTHIA. An additional potential source of
background may arise from \Pnu\Panu$\gamma$ events in which the photon
converts. This background was generated with KORALZ.

The signal was generated using SUSYGEN~\cite{bib:susygen} with final state
radiation added by the PHOTOS~\cite{bib:photos} package and tau decays
simulated with TAUOLA~\cite{bib:tauola}. A full simulation of the detector
was used for the background and the signal for some parameter sets, and a fast
simulation for interpolation of the signal efficiencies.

\section{Searches}

Slepton pair production leads to a final state characterised by 
leptons of the same flavour. 
These leptons are acoplanar with the beam due to the missing energy and 
momentum carried away by the $\chi$'s.
\par
Several searches have been developed depending on the flavour of the
leptons and the mass difference between the slepton and the $\chi$.
The main backgrounds for large mass differences are W pair production, 
four--fermion processes and dilepton production, whereas for
small mass differences the dominant background comes from two--photon
processes with lepton production. Therefore, one set of selections
has been designed for large mass differences (Section~\ref{large_dm}),
and an additional analysis was optimised for mass differences
of about 5~\Gcsq{} (Section~\ref{small_dm}) in the case of smuons and 
selectrons.

The cuts are described in this section and listed in Table~\ref{cuts}. 
The positions of the most important cuts are determined using the 
\nbarnf{} prescription advocated in~\cite{n95}, i.e., minimising the cross 
section expected to be excluded on average in the absence of a signal. 

\begin{table}[hbtp]
\begin{center}
\caption[]{\label{cuts}{Selection criteria.}}~\\
\small
\begin{tabular}{|l||l|l|l|}
\hline
& \multicolumn{2}{c|}{\bf{selectron \PSe, smuon \PSmu}} & \bf{stau \PStau} \\
& $M_{\PSl}-M_{\Pchi}>$ 6~\Gcsq{}
& $M_{\PSl}-M_{\Pchi}\leq$ 6~\Gcsq{} & \\[4pt]
\hline
\hline
charged tracks  & \multicolumn{3}{c|}{two identified leptons $(e,\mu,\tau)$} \\
\hline
neutral veto   & yes & no cut & yes \\
\hline
acollinearity   & \multicolumn{3}{c|}{$\alpha>2^{\circ}$} \\ 
\hline
acoplanarity    & \multicolumn{2}{c|}{$\Phi_{\mathrm{aco}}<170^{\circ}$} & \\
\hline
visible mass    & \multicolumn{2}{c|}{$M_{\mathrm{vis}}>$ 4~\Gcsq{}} &
                                      $M_{\mathrm{vis}}>$ 6~\Gcsq{} \\
                & & $M_{\mathrm{vis}}<20\%\sqrt{s}$ & \\
\hline
energy in $12^{\circ}$ & \multicolumn{3}{c|}{$E_{12}=0$} \\
\hline
missing         & $p_{\mathrm{T}}>3\%\sqrt{s}$ & 
                  $p_{\mathrm{T}}>1\%\sqrt{s}$ & 
               if $M_{\mathrm{vis}}<$ 30~\Gcsq{} \\
momentum      & & $p<10\%\sqrt{s}$ & then $p_{\mathrm{T}}>6\%\sqrt{s}$ \\  
              & & $|\cos{\theta}|<0.90$ & $|\cos{\theta}|<0.866$ \\
\hline
$\rho$          & $\rho>$ 2~\Gc{} & $\rho>$ 1~\Gc{} &
                  $17.1\rho+120-\Phi_{\mathrm{aco}}>0$ \\
\hline
lepton        & $ p_{1},p_{2}>0.5\%\sqrt{s}$ &
                $ p_{T1},p_{T2}>0.5\%\sqrt{s}$ & 
                $ p_{1},p_{2}>0.5\%\sqrt{s}$ \\
momenta $p_{1}, p_{2}$ & & $ p_{1},p_{2}<10\%\sqrt{s}$  & 
                  min$(p_{1},p_{2})<$ 15~\Gc{} \\
$\sqrt{s}=$161~GeV & $ p_{1},p_{2}<$ 75~\Gc{} & & 
                     $ p_{1},p_{2}<$ 30~\Gc{} \\ 
$\sqrt{s}=$172~GeV & $ p_{1},p_{2}<$ 80~\Gc{} & &
                     $ p_{1},p_{2}<$ 22~\Gc{} \\ 
\hline
$\wwchi$           & & & \\
$\sqrt{s}=$161~GeV & $\wwchi>0.5$  & & \\
$\sqrt{s}=$172~GeV & $\wwchi>2.0$  & & \\
\hline
Fisher variable  & & $y > -15$ & \\
\hline
\end{tabular}
\normalsize
\end{center}
\end{table}

The definition of a good charged track as originating from within a cylinder 
of 1~cm radius
and 10~cm length, which is centred on the nominal interaction point and 
parallel to the beam axis, having at least four TPC hits, 
a momentum greater than 
$0.5\%\sqrt{s}$ and a minimum polar angle of $18.2^\circ$ is common to all
analyses.

For selectrons and smuons, events are required to have two good charged
tracks with opposite electric charges. To search for staus, all charged
and neutral objects of an event are clustered into two jets using the
Durham algorithm, as in a large fraction of tau decays
neutral particles are produced. After having identified photon
conversions with a standard pair finding algorithm~\cite{bib:performancepaper},
one jet is required to consist of exactly one good charged track,
whereas the other is allowed to have one, two or three charged tracks.
The vector sum of the momenta of these tracks will be referred 
to as a single track for simplicity.

To avoid selecting events with a single converted photon, the 
acollinearity, defined as the angle between 
the track momenta, should be greater than $2^\circ$. 
The background coming from tagged
two--photon processes is eliminated by requiring that no energy 
be reconstructed in
a cone of $12^\circ$ around the beam axis (corresponding to an effective
threshold of 90~MeV). 
This requirement introduces an inefficiency due to beam related background
and detector noise, which
was measured to be 4$\%$ (2$\%$) at centre--of--mass energy of 161 (172)~GeV,
using events triggered at random beam crossings.

\subsection{Large Mass Differences}
\label{large_dm}

To reject radiative fermion pair production with an initial state radiation 
photon in the
detector while avoiding to veto $\tau$ decays and signal events with final
state radiation, a neutral veto is applied if three conditions are fulfilled
simultaneously: a neutral energy flow object of more than 4~GeV 
is reconstructed, its
angle with each of 
the two tracks is greater than $10^\circ$ and its
invariant mass with each of the two tracks is greater than 2~\Gcsq{}.

\subsubsection{Selectrons and Smuons}
\label{standard}

In order to reject events from fermion pair production, the acoplanarity
angle $\Phi_{\mathrm{aco}}$ of the two tracks is required to be less than 
170$^\circ$. 
The acoplanarity is defined as the angle between the track momenta projected
onto a plane perpendicular to the beam axis.
To reject the remaining ee$\gamma$ events, the energy of the 
tracks, including neutral objects in a cone of 10$^\circ$ around either
track, is required
to be less than 75~(80)~GeV at centre--of--mass energy of 161~(172)~GeV.
\par
\begin{figure}
\vspace{-3.cm}
\epsfig{file=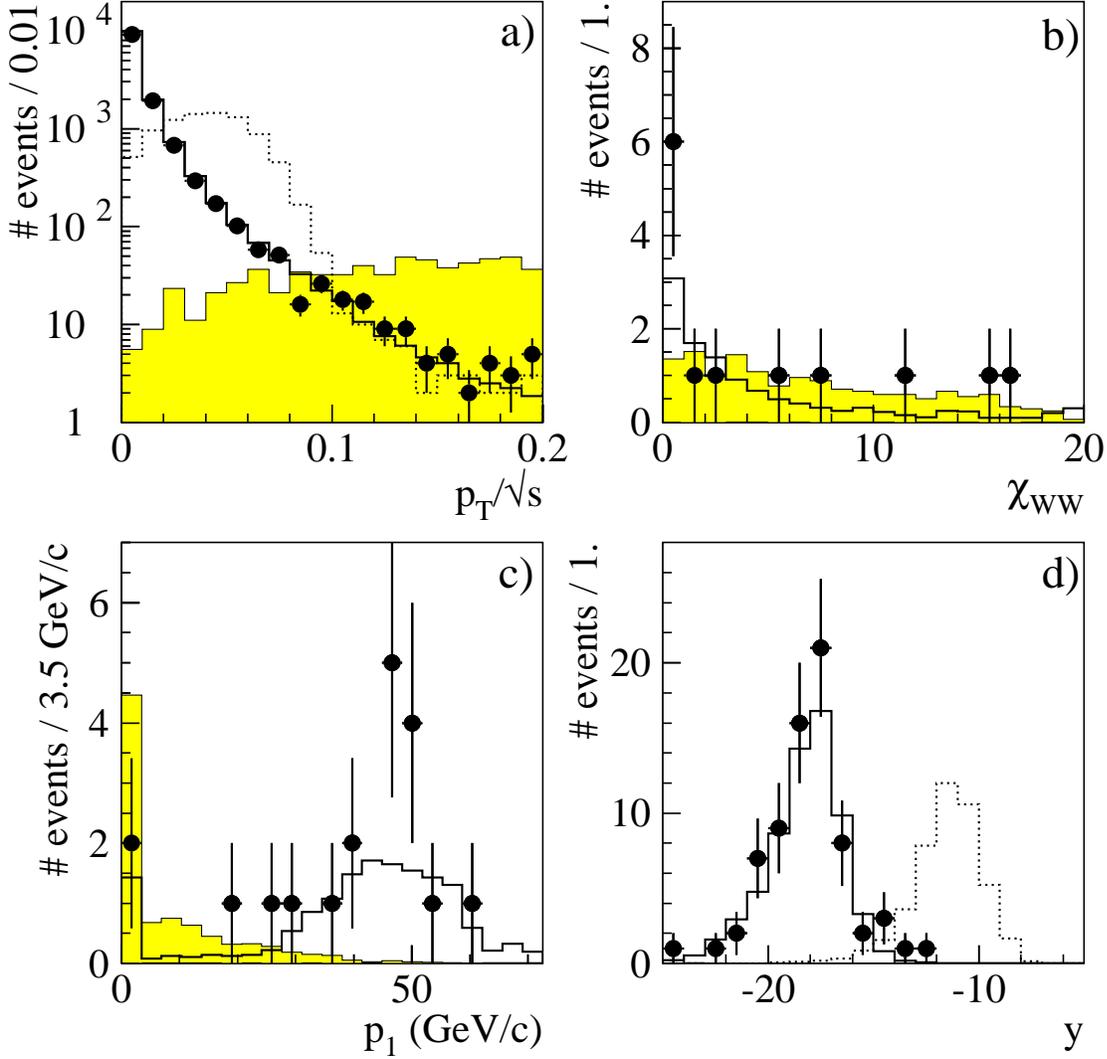,width=16cm}
\caption[]{\label{data}{Distributions of 
  a) the missing transverse momentum, 
  b) the \wwchi{} variable for the selectrons and smuons, 
  c) the momentum of the leading lepton $p_{1}$ as used in the stau search, 
  d) the Fisher variable. 
  The points are the data (161~GeV and 172~GeV combined), the open
  histograms the background Monte Carlo normalised to the recorded 
  luminosity, the shaded histograms a signal with a mass difference
  of 30~\Gcsq{} in arbitrary normalisation, and the dotted histograms 
  in plots a) and d) are signal histograms with a mass difference of 
  5~\Gcsq{}. In order to preserve sufficient statistics only subsets
  of the cuts on the other variables were applied for these plots.}}
\end{figure}

The (non--WW) four--fermion and two--photon backgrounds are reduced by
demanding that the visible mass be greater than 4~\Gcsq{}.
The untagged two--photon processes are reduced further by demanding that
the missing transverse momentum $p_T$ of the event (Fig.~\ref{data}a)
be greater than $3\%\sqrt{s}$.
Remaining two--photon events ($\gamma\gamma\rightarrow\tau\tau$)
are reduced by the following procedure: 
the track momenta are projected onto a plane transverse to the beam axis
and the thrust axis is calculated from the projected momenta. 
The scalar sum~$\rho$ of
the transverse components of the projected momenta with respect 
to this thrust axis is required
to be greater than 2~\Gc{}.
\par
At this point of the analysis, only the background from W pair production,
where both W's decay leptonically,
has to be dealt with. In only two out of nine cases the W's are expected to 
decay to two electrons or two muons.
Therefore two identified electrons (muons) are required for the selectron
(smuon) search.
\par
The cuts described so far are common to both centre--of--mass energies.
The cross section for W pair production, however, increases almost by
a factor of four going from 161~GeV to 172~GeV.
Therefore more stringent cuts are applied in the latter case.
\par
For leptons from W pair production, a lepton energy of roughly $\sqrt{s}/4$
smeared with the boost and width of the W is expected. Therefore the variable
\begin{displaymath}
  \label{eq:wwchi}
  \wwchi = \frac{1}{2}\sum_{i=1}^2 
  \left(\frac{E_i-\sqrt{s}/4}{6\;\mathrm{GeV}}\right)^2
\end{displaymath}
is defined, where $E_i$ are the lepton energies and 6~GeV is the expected 
energy spread. For the W background this variable peaks
near zero while for a signal not too similar to a W it is expected to be
flatter (Fig.~\ref{data}b).
At 161~GeV, $\wwchi>0.5$ is required. This cut is tightened to $\wwchi>2$
at 172~GeV to cope with the
higher W cross section and the broader distribution of the lepton momenta.
It is not applied for $M_{\PSl}>70$~\Gcsq{} to save efficiency. The
point of the transition was determined with the \nbarnf{} procedure.
After these cuts the residual WW background is still much higher at 172~GeV
than at 161~GeV. In order to reduce the background even further at 172~GeV,
it is required that the lepton momenta must fall in the range
kinematically allowed for the specific combination of slepton and $\chi$ 
masses.
Typical efficiencies and background expectations for $\PSl_R\PSl_R$ 
production
for these searches are listed in Table~\ref{eff}.

\subsubsection{Staus}
\label{stau}
Background from untagged two--photon processes is efficiently suppressed by
requiring a missing transverse momentum of at least $6\%\sqrt{s}$, if the
visible mass of the event is less than 30~\Gcsq{}. 
Events from fermion pair production and two--photon processes are rejected by
a two--dimensional cut in $\rho$ and
$\Phi_{\mathrm{aco}}$ (defined as before, but using jets instead of
track momenta): ($\rho$,$\Phi_{\mathrm{aco}}$) is not allowed in the triangular
region ($0,120$) - ($3.5,180$) - ($0,180$).
Furthermore the jet masses are required to be less than 8~\Gcsq{}. Remaining
difermion events with an ISR photon undetected in the beam pipe are suppressed
by requiring a polar angle of the missing momentum of at least $30^\circ$.

After requiring a minimum visible mass of 6~\Gcsq{}, the remaining background
predominantly consists of events from W pair production. As the leptons from
W bosons decaying into electrons or muons are in general more energetic
than leptons from $\tau$ decays, cuts are applied on the momenta of
identified electrons or muons (including all neutral objects within 
$10^\circ$). 
At $\sqrt{s}=161$ (172)~GeV, the leading lepton 
is required to have a momentum $p_{1}$ less 
than 30 (22)~\Gc{} (Fig.~\ref{data}c).
In case a second lepton is found in the event, its momentum is required to
be less than 15~\Gc{}.

After these cuts a total background of 0.5 (1) events is expected at
161 (172)~GeV. Examples of the efficiencies to select stau events are
presented in Table~\ref{eff}.

\subsection{Small Mass Differences}
\label{small_dm}
\begin{table}[htb]
\vspace{-1cm}
\begin{center}
\caption[]{\label{eff}{Signal cross sections $\sigma_{\PSl_{R}}$
($\tb=2$, $\mu=-200~\Gcsq{}$ for $\PSe_{R}$), efficiencies $\epsilon$ 
and background cross sections $\sigma_{B}$ for slepton searches. 
For simplicity, the background is given for each mass combination 
separately in spite of the overlap among the various rows.}}~\\
\begin{tabular}{|c|c|c|c|c|c|c|}
\hline
 $\sqrt{s}$ & Slepton &  $M_{\PSl}$  & $M_{\chi}$ & 
\hspace{0.5cm}$\sigma_{\PSl_R}$\hspace{0.5cm} & 
\hspace{0.7cm}$\epsilon$\hspace{0.7cm} & 
\hspace{0.5cm}$\sigma_{B}$\hspace{0.5cm} \\[4pt]
   (GeV)      &      &  (\Gcsq{})   & (\Gcsq{})   &   (fb)  & (\%)   &   (fb)     \\
\hline
\hline
            &         &     75       & \dum{}0     &    238   &   58      &    60     \\ 
            &         &     75       &     30      &    160   &   58      &    60     \\
            &\rb{\PSe}&     75       &     70      &\dum{}38  &   45      &    35     \\
            &         &     75       & \pdum{}72.5 &\dum{}34 &\dum{}9    &    35     \\ \cline{2-7}
            &         &     55       & \dum{}0     &          &   66      &    43     \\
\rb{161}    &         &     55       &     30      &          &   61      &    43     \\
            &\rb{\PSmu}&    55       &     50      & \rb{401} &   54      &    35     \\ 
            &         &     55       & \pdum{}52.5 &          &   18      &    35     \\ \cline{2-7}
            &         &     50       & \dum{}0     &          &   39      &    43     \\
          &\rb{\PStau}&     50       &     25      & \rb{514} &   40      &    43     \\
\hline
            &         &     75       & \dum{}0     &    619   &   67      & 117\dum{} \\
            &         &     75       &     30      &    415   &   67      &    91     \\
            &\rb{\PSe}&     75       &     70      &    107   &   45      &    35     \\
            &         &     75       & \pdum{}72.5 &\dum{}97  &\dum{}9    &    35     \\ \cline{2-7}
            &         &     55       & \dum{}0     &          &   60      &    93     \\
\rb{172}    &         &     55       &     30      &          &   55      &    54     \\
            &\rb{\PSmu}&    55       &     50      & \rb{411} &   54      &    35     \\
            &         &     55       & \pdum{}52.5 &          &   18      &    35     \\ \cline{2-7}
            &         &     50       & \dum{}0     &          &   37      &    93     \\
           &\rb{\PStau}&    50       &     25      & \rb{502 }&   36      &    93     \\
\hline
\end{tabular}
\end{center}
\end{table}

The analysis described in this section
is optimised for small mass differences. It is used 
for mass differences less than 6~\Gcsq{}.
The background from fermion pair production and four--fermion processes is
rejected by demanding a maximum lepton momentum of $10\%\sqrt{s}$, a visible
mass smaller than $20\%\sqrt{s}$, a missing momentum of the event smaller
than $10\%\sqrt{s}$ and the acoplanarity angle of the two leptons to be 
below 170$^\circ$.

Since the signal resembles the two--photon background, the cuts are less 
stringent than in the large mass difference selection.
The missing 
transverse momentum is required to exceed $1\%\sqrt{s}$ 
(Fig.~\ref{data}a),
$\rho$ is required to be greater than 1~\Gc{} and
the visible mass is required to be greater than 4~\Gcsq{}.
The cosine of the polar angle of the missing momentum must be less than
0.9 and
the transverse momentum of each lepton greater than
$0.5\%\sqrt{s}$.

In order to reduce the large \gaga{}--background a Fisher discriminant
analysis \cite{fisher} has been used. 
This method exploits the remaining modest differences between \gaga{}--events
and the signal, taking into account the correlations among the variables used.
For this analysis, the visible event mass, the missing transverse momentum,
the missing momentum along the beam direction, the variable $\rho$, 
the maximum and minimum
lepton transverse momenta and the acollinearity are used to calculate the
Fisher variable $y$ (Fig.~\ref{data}d).
For $y > -15$, the cut chosen by means of the \nbarnf{} procedure, 
$98\%$ of the remaining background is rejected, whereas about $98\%$ of a
signal with a mass difference of 5~\Gcsq{} is kept.
After these cuts a total background of 1 (1) event is expected in the whole 
data sample for selectrons (smuons).

\section{Results}

Three candidates 
are observed in the large mass difference analyses and are listed in
Table~\ref{cand}.
The selectron candidate at 161~GeV can be either interpreted as a WW event,
where the first electron originates from the W decay and the second
from a cascade decay W to $\tau$ to electron, or as a \PZz$\gamma^{\ast}$
event with the \PZz\ decaying to neutrinos, since the recoil mass to the 
electron pair is 90~\Gcsq{}.
The decay products of the stau candidate, observed at 161~GeV, are consistent
with the masses of the $\rho$ and $a_1$ mesons. 
From the decay kinematics
an upper limit on the tau energy of 22~GeV can be inferred. 
Therefore it is unlikely that this is a WW event, but it is compatible with 
\PZz$\gamma^{\ast}$.
One smuon candidate is observed at 172~GeV. The recoil mass is 
130~\Gcsq{}, so that
the WW (cascade decays to muons via taus) 
and the \PZz$\gamma^{\ast}$ hypotheses are possible explanations.
\par
In the analysis optimised for small mass difference five candidates compatible
with \gaga{} and WW production are
observed and listed also in Table~\ref{cand}.

\begin{table}[htbp]
\begin{center}
\caption[]{\label{cand}{Kinematic properties of the candidate events.}}
~\\
\begin{tabular}{|c|c|c|c|c|c|}
\hline
$\sqrt{s}$ & Slepton & $M_{\PSl}-M_{\chi}$ &  $p_{1}$  & $p_{2}$  
& event \pT{} \\[4pt]
  (GeV)      &         & (\Gcsq{})            &  (\Gc{})   & (\Gc{})   
& ~\Gc{}      \\
\hline
\hline
            & \PSe    &              & 44.1 & 12.0 &    23.1    \\ 
\cline{2-2}
\cline{4-6}
\rb{161}    & \PStau  & $>6$         & 11.2 & \dum{}9.7 &      10.6 \\ 
\cline{1-2}
\cline{4-6}
  172       & \PSmu   &              & 19.4 & 17.4 &   22.8        \\
\hline
\hline
  161       & \PSmu   &              & \dum{}3.9 & \dum{}2.4 & \dum{}3.1 \\ 
\cline{1-2}
\cline{4-6}
           &         &               & \dum{}6.2 & \dum{}2.7 & \dum{}4.8 \\
           &\rb{\PSmu}& $\leq 6$     & \dum{}8.0 & \dum{}1.9 & \dum{}4.5 \\
\cline{2-2}
\cline{4-6}
\rb{172}    &         &              & 16.7      & \dum{}1.8 & 10.7      \\
            &\rb{\PSe}&              & \dum{}6.9 & \dum{}5.9 & \dum{}2.6 \\
\hline
\end{tabular}
\end{center}
\end{table}

To summarise, in the $(M_{\PSl},M_{\chi})$ plane, for $M_{\PSl}$ greater than
45~\Gcsq{}, eight events are selected in the data in agreement
with the seven events expected from standard model processes.
In the absence of a signal, limits are set for various processes.

\subsection*{Limits}
The limits are derived from the present results 
combined with the ALEPH results from LEP1.5~\cite{bib:lep15paper},
obtained at $\sqrt{s}=130-136~\Gcsq{}$. 
For each combination of slepton and neutralino masses only 
the candidates
that fulfil the kinematic requirements of the specific combination are taken
into account in the calculation of the limit.
Systematic uncertainties are taken into account by reducing the 
expected number of signal events by one standard deviation. 
The main contributions to the total systematic error ($\sim 3\%$) come from
Monte Carlo statistics, the luminosity measurement ($<1\%$) 
and lepton identification
efficiencies ($\sim 2\%$).

\begin{figure}
\parbox{7.5cm}{\epsfig{file=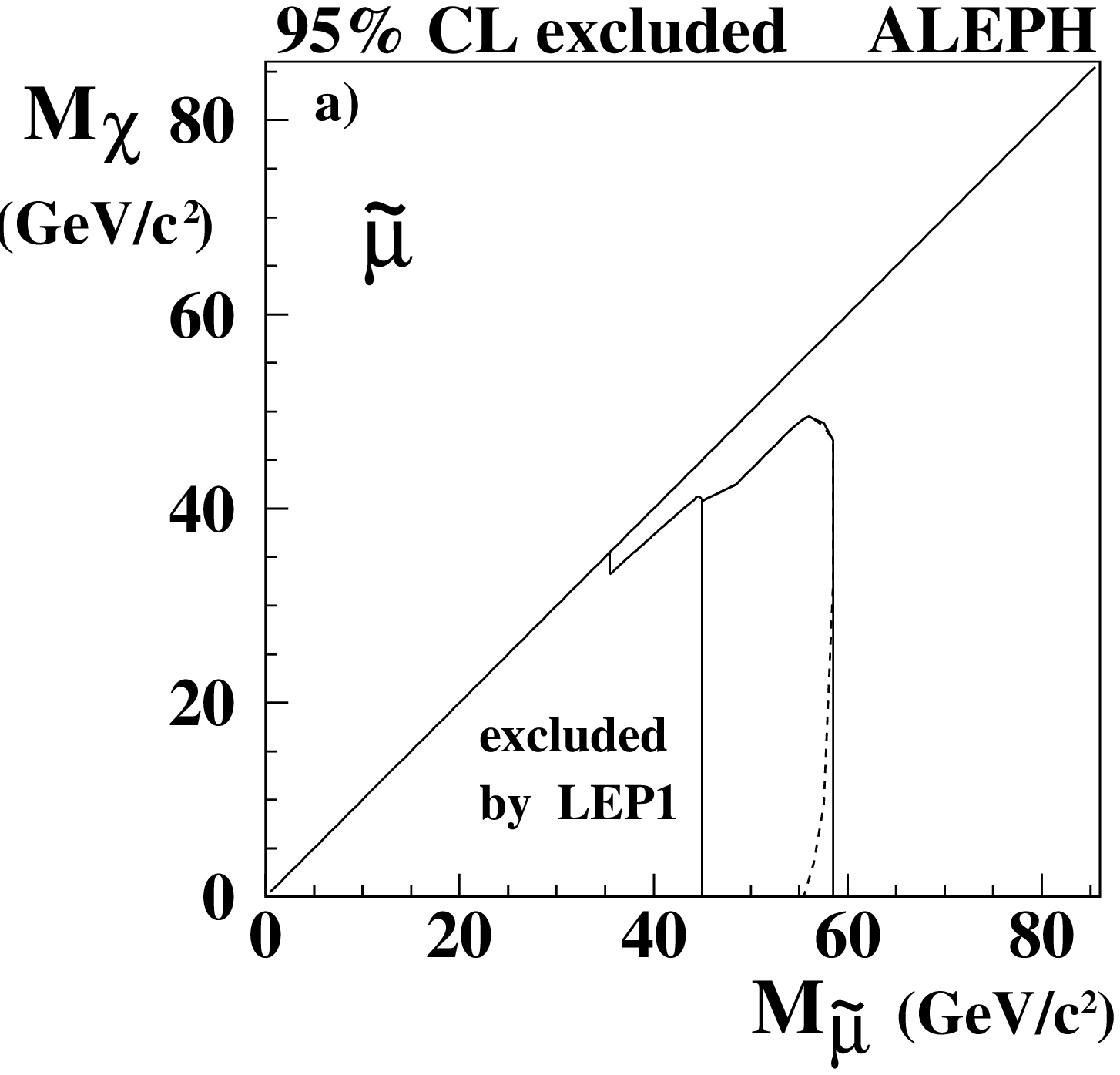,width=8.5cm}}
\hfill
\parbox{8cm}{\epsfig{file=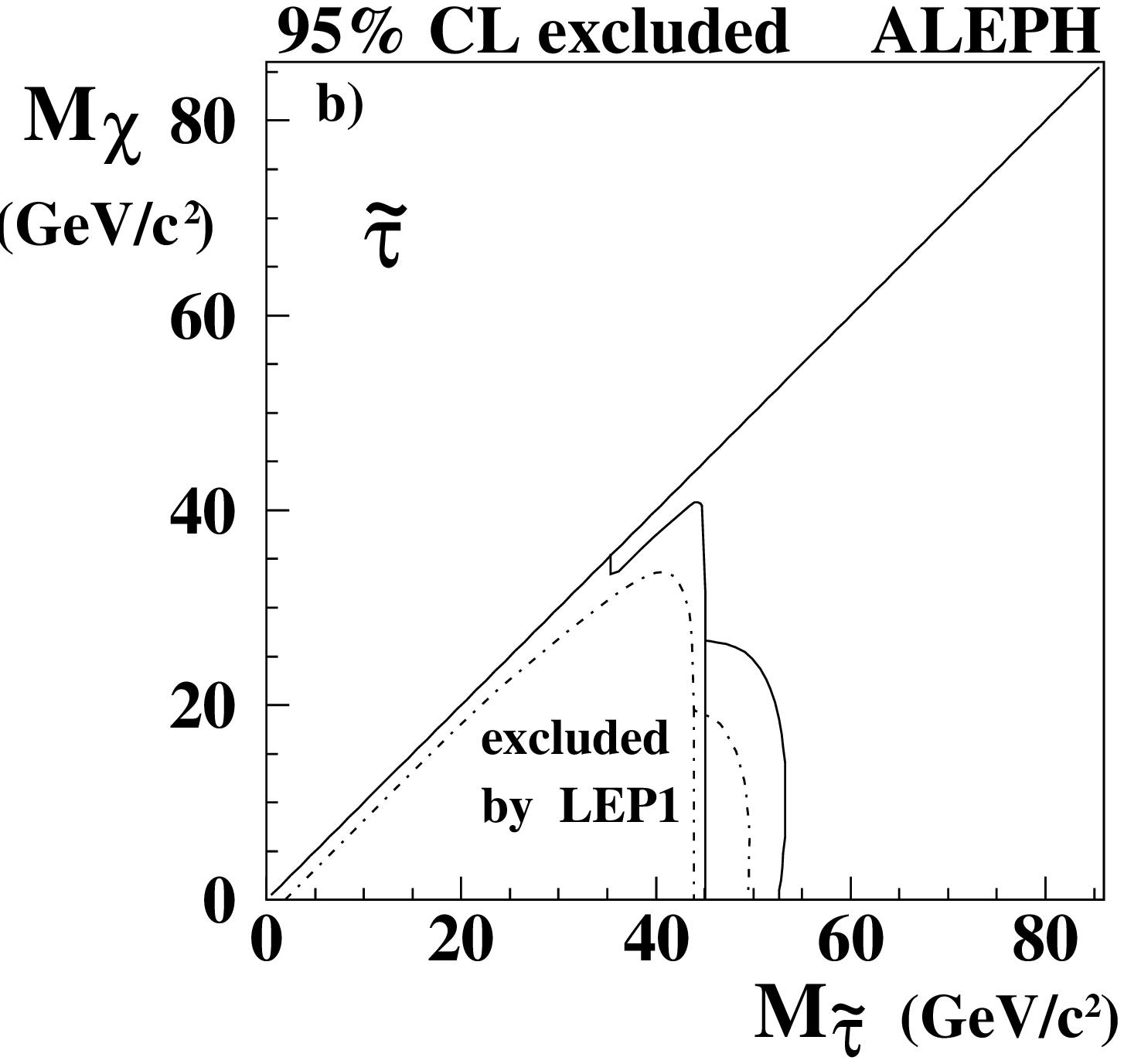,width=8.5cm}}
\parbox{7.5cm}{\epsfig{file=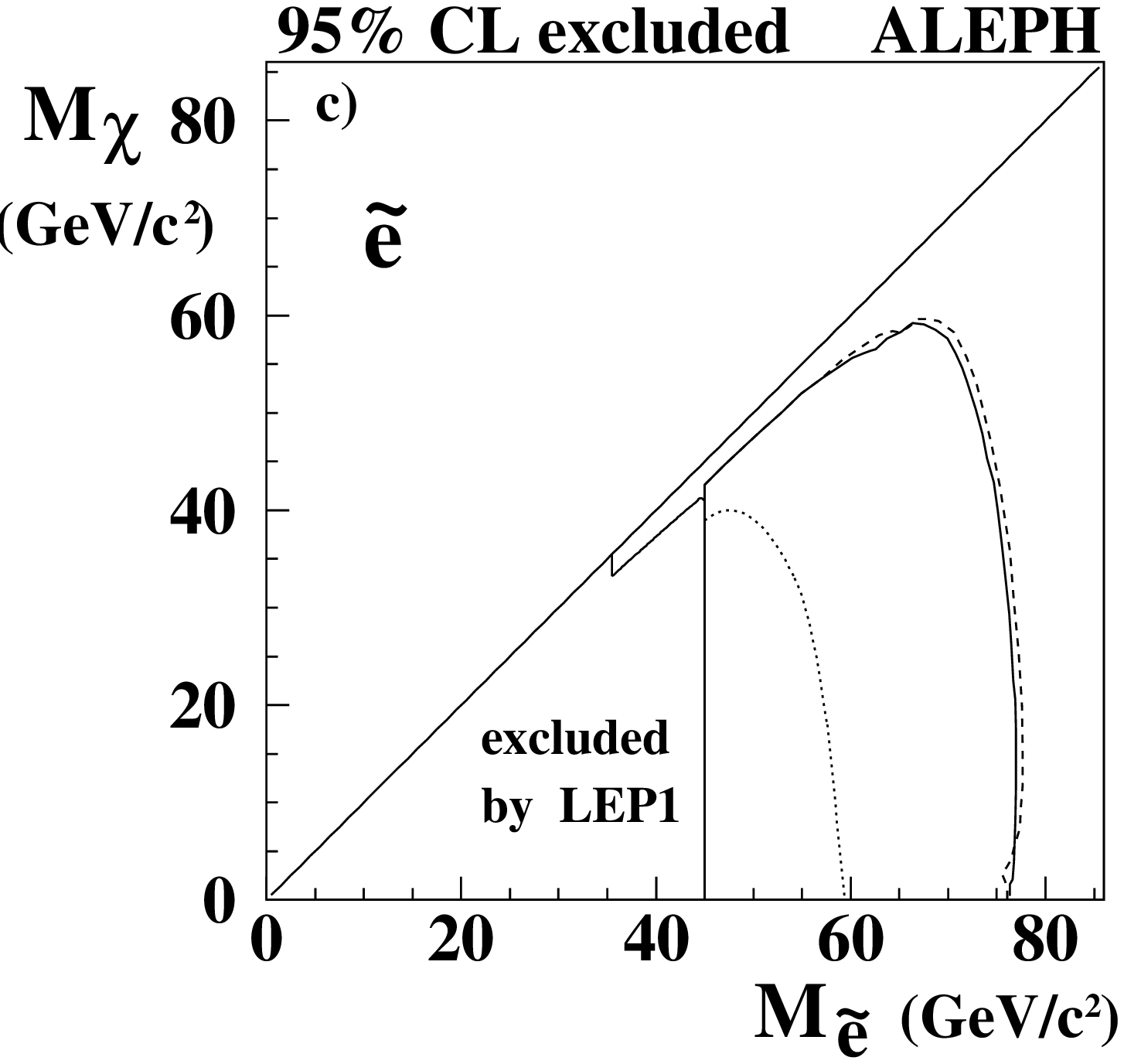,width=8.5cm}}
\hfill
\parbox{8cm}{\epsfig{file=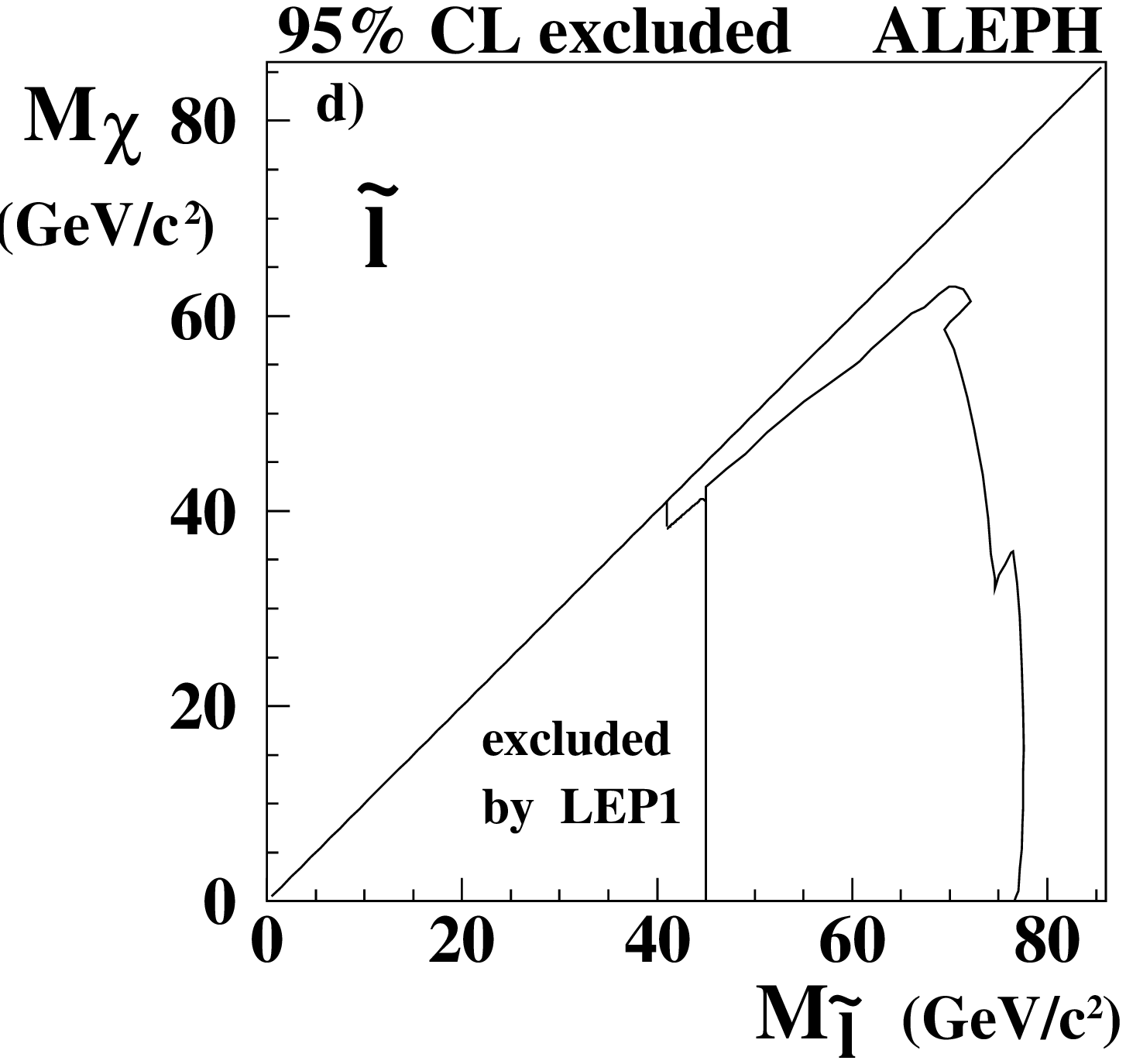,width=8.5cm}}

\caption[]{\label{mass_lim}{a) The solid curve shows the limit obtained for
  $\PSmu_R$ assuming $\mathrm{BR}(\PSmu_R \rightarrow \mu \chi) = 100\%$ 
  and the dashed curve shows the effect of cascade decays for
  $\mu=-200~\Gcsq$ and $\tb=2$, assuming no efficiency for the cascade decays.
  b) Mass limits for $\PStau_R$ (solid curve) and a mixed state $\PStau_1$
  decoupled from the Z (dash--dotted curve), assuming 
  $\mathrm{BR}(\PStau \rightarrow \tau \chi)=100\%$. 
  c) $\PSe_R$ mass limit for $\tb=2$. The solid curve shows the limit for the
  case $\mu=-200~\Gcsq{}$, the dashed curve for $\mu=1000~\Gcsq{}$ assuming
  no efficiency for cascade decays. The dotted curve is the LEP~1.5 limit. 
  d) $\PSe_R$ and $\PSmu_R$ limits, combined assuming mass degeneracy and 
  including the effect of cascade decays for $\tb=2$ and $\mu=-200~\Gcsq{}$.}}
\end{figure}

For smuons and staus, fewer assumptions are necessary to derive a mass
limit than for selectrons, for which the \tchannel{} production requires
specification of the neutralino parameters.
Unless stated otherwise, limits are derived under the 
assumption that only $\PSl_R\PSl_R$ production contributes. 
This assumption is conservative because of the smaller cross section for the
production of right-handed sleptons compared to left-handed sleptons
for pure \schannel{} production.

Using the limit on additional contributions to the invisible width of
the \PZz{} as derived with the full LEP1 statistics \cite{width} (sleptons
do not contribute to the leptonic width, since they fail the cut
on the acollinearity in the standard analysis),
lower limits on slepton masses can be set independent of the neutralino
mass. For a single right-handed slepton this corresponds to 35~\Gcsq{}.
Assuming all three right-handed sleptons to be degenerate in mass,
slepton masses below 41~\Gcsq{} are excluded.
\par
The limit on the \PSmu$_R$ is shown in Figure~\ref{mass_lim}a.
Smuon masses up to 59~\Gcsq{} are excluded 
for mass differences to the lightest neutralino greater than 10~\Gcsq{}. 
This limit is reduced when cascade decays
via heavier neutralinos
are taken into account.
Conservatively no 
efficiency is assumed for these decays.
The dashed curve shows this effect for $\mu=-200~\Gcsq$ and $\tb = 2$
as an example.

Mixing is expected to be negligible for all sleptons except for staus, since
the tau is much heavier than the other leptons. Therefore
limits are calculated for staus in mixed and unmixed scenarios.
Assuming mixing effects to be 
negligible, the most conservative limits are set by
considering pair production of $\PStau_R$ (Fig.~\ref{mass_lim}b, solid
curve).
In the case where $\PStau_L$ and $\PStau_R$ mix, 
limits are set on the mass 
of the lightest stau $\PStau_1$, choosing the mixing angle such that $
\PStau_1$ completely decouples from the Z boson
(Fig.~\ref{mass_lim}b, dash--dotted curve). 
For this purpose the search for staus at LEP1 \cite{lep1} is
updated with the full LEP1 statistics and included in the determination
of the limit.
For a mass difference between stau and
neutralino of more than 30~\Gcsq{}, the $\PStau_R$ ($\PStau_1$)
with mass less than 53~\Gcsq{} (47~\Gcsq{}) is excluded.

Limits for the selectrons are obtained in the gaugino region, where 
the cross section is enhanced due to the \tchannel{} contribution.
The limits are shown in Fig.~\ref{mass_lim}c using two different values of
$\mu$ (1000~\Gcsq{} and $-200$~\Gcsq{}) for $\tb=2$. 
The accessible region for the candidate of the large mass difference
selection and one of the candidates of the small mass difference selection is
entirely excluded.
The actual limit is in a region of the ($M_{\PSe},M_{\Pchi}$) plane where
only one candidate in the small mass difference domain must be considered.
Cascade decays are taken into account according to the branching ratio
$\PSe\to\Pe\Pchi$ for the particular choice of the MSSM parameters, assuming
no efficiency for cascades. 
This leads to a degradation of the selectron mass limit for small
$M_{\Pchi}$.
The effect of cascade decays is not as pronounced as for the smuons due
to the increasing cross section for small neutralino masses.
The limit shown in Fig.~\ref{mass_lim}c is extended by the single photon
counting measurements
to 80~\Gcsq{} for neutralino masses
less than 10~\Gcsq{} \cite{singlephot} under the assumption of
degenerate left and right selectrons at 90\% confidence level. 

A limit on $M_{\PSl}$ is derived assuming degeneracy of the three 
flavours. The highest sensitivity in the direct
search is reached when only selectrons and smuons are combined, since 
staus are selected with similar background but lower efficiency.
The result for $\tb=2$ and $\mu=-200~\Gcsq{}$ is shown
in Fig.~\ref{mass_lim}d. Only $\PSl_R\PSl_R$ production is
considered. The smuon candidate of the large mass difference selection causes
a notch in the limit curve.

\begin{figure}
\vspace{-2cm}
\epsfig{file=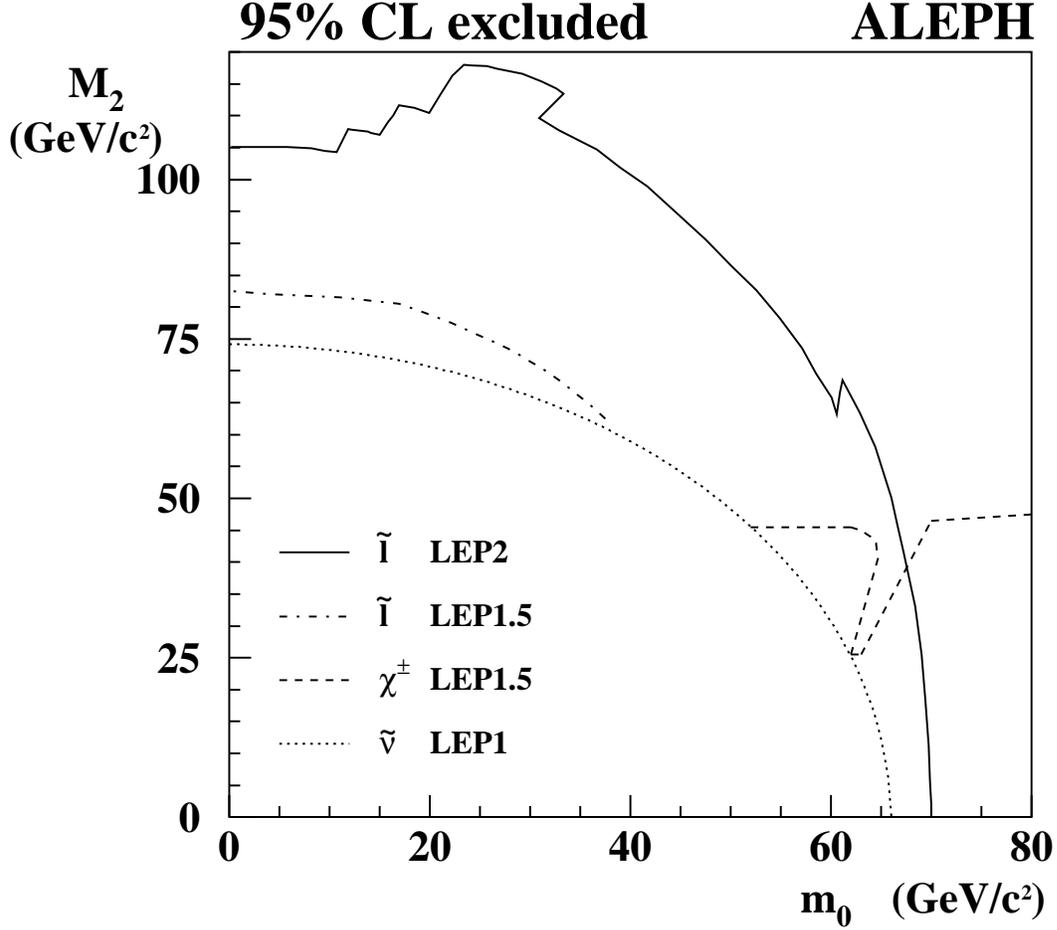,width=16cm}
\caption[]{\label{gut_lim}{Limit in the ($m_0,M_2$) plane combining selectrons
  and smuons for $\tb=2$ and $\mu=-200~\Gcsq$ (solid curve). 
  The dotted curve shows the sneutrino limit (43~\Gcsq{}) from LEP1,
  the dashed curve shows the gain in exclusion due to the chargino limit
  from LEP1.5 and the dash--dotted curve shows the slepton limit of LEP1.5.}} 
\end{figure}

Assuming scalar mass unification ($m_0$) at the GUT scale there is a relation
among the masses of the scalar particles at the electroweak scale~\cite{m0}.
In particular, for sleptons one obtains:
\begin{eqnarray*}
  M_{\PSl_R}^2 &=& m_0^2 +0.22M_2^2-\sqtw M_{\rm Z}^2\cos2\beta\\
  M_{\PSl_L}^2 &=& m_0^2 +0.75M_2^2
  -0.5(1-2\sqtw)M_{\rm Z}^2\cos2\beta\\
  M_{\PSnu}^2 &=& m_0^2 +0.75M_2^2+0.5 M_{\rm Z}^2\cos2\beta
\end{eqnarray*}
In the case of the staus, mixing plays a role for large values of \tb{}
and/or $\mu$. Depending on the masses of left- and right-handed sleptons,
additional production channels like $\PSe_R\PSe_L$ and $\PSl_L\PSl_L$ may be
open leading to a higher total cross section. Again, 
the highest sensitivity is obtained when only
selectrons and smuons are used for the combination. The excluded region in
the ($m_0,M_2$) plane for $\tb=2$ and $\mu=-200~\Gcsq{}$ is shown in
Fig.~\ref{gut_lim}. The notch for intermediate values of $m_0$ comes from the
smuon candidate of the large mass difference selection, and the structures
for small $m_0$ come from the various small mass difference candidates.
For small $M_2$, i.e., small neutralino masses,
the processes $\PSl_L\PSl_L$ and $\PSe_L\PSe_R$ improve
only by about $1~\Gcsq{}$
the limit of
degenerate $\PSl_R$ 
because of the high branching ratio of
$\PSl_L$ to charginos, which dominantly decay hadronically in this
region.
While the limit on the sneutrino mass at LEP~1 still improved
the limit at LEP~1.5, this is not the case anymore.

\section{Conclusions}

In data samples of 11.1~\pbinv{} and 10.7~\pbinv{} recorded in 1996 by
the ALEPH detector at LEP at centre--of--mass energies of 161~GeV and
172~GeV, searches for signals of scalar lepton production have been performed. 
The number of candidate events observed is consistent with 
the background expected from
four--fermion processes and \gaga{}--interactions.
The following limits have been set at $95\%$ confidence level
($\mu=-200$~\Gcsq{} and $\tb=2$, where relevant):
\renewcommand{\labelitemi}{--}
\begin{itemize}
\item 59~\Gcsq{} for right smuons when $M_{\PSmu_R}-M_{\chi}$ 
  is greater than 10~\Gcsq{},
\item 53~\Gcsq{} for right staus if $M_{\chi}$ is smaller than 20~\Gcsq{},
\item 75~\Gcsq{} for selectrons when $M_{\PSe_R}-M_{\chi}$
      is greater than 35~\Gcsq{}, taking cascade decays into account,
\item 58~\Gcsq{} for selectrons with $M_{\PSe_R}-M_{\chi}$
   at least 3~\Gcsq{},
\item 76~\Gcsq{} for mass degenerate sleptons if $M_{\chi}$ 
   is smaller than 30~\Gcsq{}, taking cascade decays into account.
\end{itemize}
The limit in the ($m_0,M_2$) plane under the 
assumption of scalar mass unification
at the GUT scale is shown in Fig.~\ref{gut_lim}.
These results substantially extend the domains previously excluded at LEP.

\section*{Acknowledgements}

It is a pleasure to congratulate our colleagues from the accelerator divisions
for the successful operation of LEP above the W threshold. 
We would like to express our gratitude to the engineers and 
support people at our home institutes without whose dedicated help
this work would not have been possible. 
Those of us from non--member states thank CERN for its hospitality
and support.

\end{document}